\documentclass[epj]{svjour}
\usepackage{latexsym}
\usepackage{graphics}
\begin{document}
\title{Probing dense and hot matter with low-mass dileptons and photons}
\author{
   Itzhak Tserruya
   \thanks{\emph {Work supported by the Israeli Science Foundation}}
}

\institute{Weizmann Institute of Science, Rehovot 76100, Israel }
\date{Received: date / Revised version: date}
%
\abstract{Results on low-mass dileptons, covering the very broad energy range from the BEVALAC up to SPS are reviewed. 
The emphasis is on the open questions raised by the intriguing results obtained so far and the prospects for addressing 
them in the near future with the second generation of experiments, in particular HADES, NA60 and PHENIX.
   \PACS{
      {PACS-key}{discribing text of that key}   \and
      {PACS-key}{discribing text of that key}
     } 
} 
\maketitle
\section{Introduction}
\label{intro}
Electromagnetic probes are very valuable to diagnose the hot and dense matter produced in ultra-relativistic
heavy-ion collisions. They play a crucial role in the quest for the QCD phase transition to the Quark-Gluon Plasma, the
state of matter predicted by lattice QCD numerical calculations \cite{lattice} and characterized by deconfinement of 
quarks and gluons and
restoration of chiral symmetry. Due to their large mean-free-path, these probes do not suffer from final state
interactions and once produced they can escape unaffected to the detector carrying information about the conditions and
properties of the medium at the time of production \cite{shuryak78}. The interest is in the detection of dileptons and
photons emitted early in the collision which can carry direct evidence of deconfinement or chiral symmetry restoration (CSR).
The challenge is to identify them among the overwhelming yield of dileptons and photons which are produced later in the
collision, after freeze-out, from hadron decays. Precision p+p and p+A data are therefore an essential pre-requisite for
a detailed mapping of these hadronic sources.

\begin{figure}
\resizebox{0.50\textwidth}{!} {\includegraphics{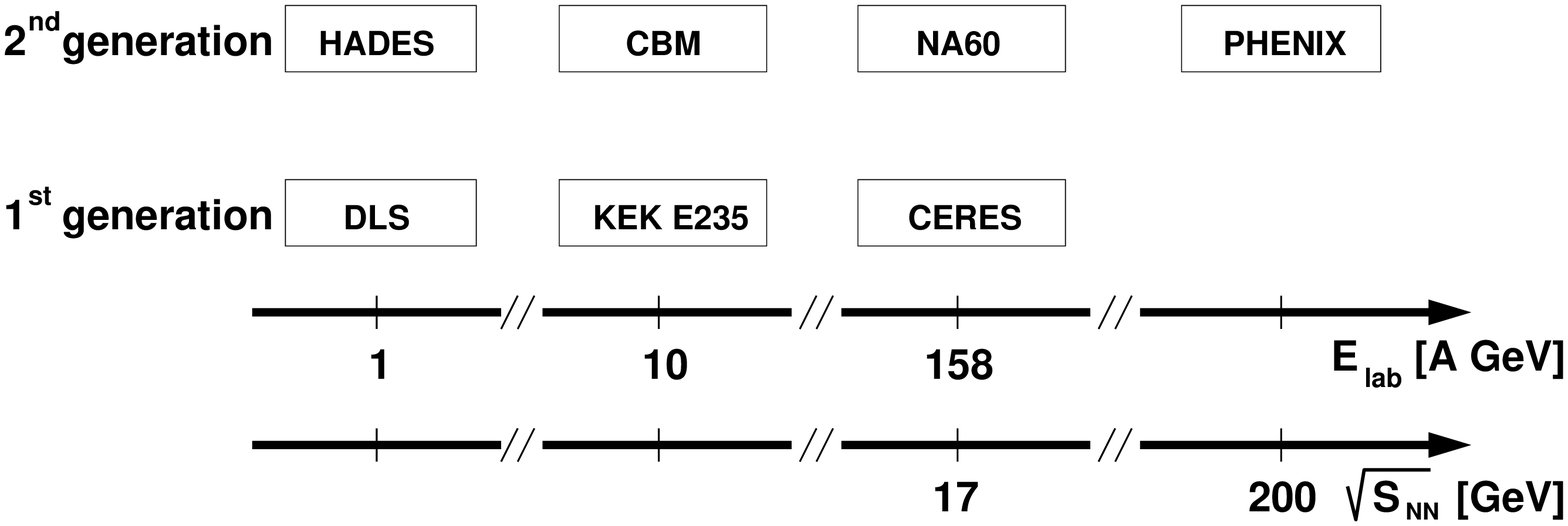}}
\caption{Low-mass dilepton experiments at a glance}
\label{fig:exp}
\end{figure}

The physics potential of low-mass dileptons (m$_{e^+e^-} \leq $1 GeV/c$^2$) has been confirmed by the interesting and intriguing 
results obtained so far and by the large number of new experiments focusing on their study. The main experiments 
are presented in Fig.~\ref{fig:exp}. One can distinguish two generations. The first generation consists of 
experiments already completed. They covered a rather broad energy range: the DLS at the BEVALAC studied dielectron production 
at 1 GeV per nucleon \cite{dls97}, the E325 experiment
at the KEK-PS studied vector meson production in cold nuclear matter using p+C and p+Cu collisions at 12~GeV \cite{kek} and the CERES 
experiment at the CERN SPS  studied nuclear collisions from 40 up to 200~GeV per nucleon \cite{ceres-p,ceres-s,ceres-pb158,ceres-pb40}. 
For completeness one should also mention the HELIOS-3 \cite{helios3} and the NA38/50 \cite{na50-low-mass} experiments which also produced
results on low-mass dileptons. The second generation consists of 4 experiments, three of them, HADES, CBM, and NA60
repeat or extend under different or improved conditions the experiments (DLS, E325 and CERES, respectively) of the
first generation. The fourth experiment of the second generation, PHENIX, opens new ground by exploring electromagnetic
probes at RHIC energies.

The discussion in this paper is limited to three topics. In Section 2, I review the most
interesting results on low-mass dileptons, and their possible link to chiral symmetry restoration, from the first
generation of experiments emphasizing the open or controversial issues which are expected to be answered by the
experiments of the second generation. In Section 3, I summarize the controversial results on the $\phi$ meson
production at the CERN SPS and in Section 4 I examine the search for thermal radiation.  A short summary is given in
Section 5.

\section{Low-mass dileptons}
\label{sec:low-mass}
\subsection{Results from CERN SPS}
\label{subsec:SPS}
\begin{figure}
  \resizebox{0.50\textwidth}{!} {\includegraphics{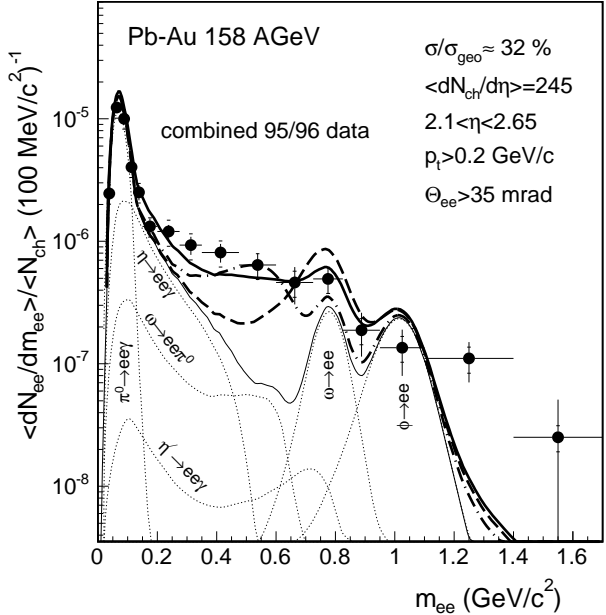} }
  \caption{CERES invariant mass spectrum of
           $e^+e^-$ pairs compared to the expected cocktail of hadron decays (thin solid line). The figure
           includes also calculations assuming the vacuum $\rho$ spectral shape (dashed), $\rho$ dropping mass (dash-dotted) and
           in-medium $\rho$ broadening (solid) \cite{ceres-95-96}. }
\label{fig:ceres-95-96}
\end{figure}

The low-mass pair continuum has been systematically studied by the CERES experiment at CERN in p+Be (a good
approximation to p+p), p+Au \cite{ceres-p}, S+Au \cite{ceres-s} and Pb+Au collisions \cite{ceres-pb158,ceres-pb40}. The results
confirm the unique physics potential of this probe. The most prominent feature is the enhancement of electron pairs
observed in the mass region m=0.2-0.6 GeV/c$^2$ in all heavy-ion collisions studied whereas in p+Be and p+Au collisions
the spectrum is well described by the known hadron decays. 
A recent example of the enhancement observed by CERES in Pb+Au
collisions at 158 A GeV  from the combined analysis of the 95 and 96 runs is shown in 
Fig.~\ref{fig:ceres-95-96} \cite{ceres-95-96}. The results are presented in the form of pair 
production per event and per charged particle within
the spectrometer acceptance and are compared to the cocktail of known hadronic sources (thin solid line). The yield is
clearly enhanced with respect to this cocktail. CERES has also shown that this excess is mainly due to soft p$_T$ pairs
and that it increases faster than linearly with the event multiplicity \cite{ceres-95-96}.

This low-mass dilepton enhancement is one of the highlights of the 15 year heavy-ion program at CERN. Its possible
connection to CSR has triggered a wealth of theoretical activity (for recent comprehensive reviews see
\cite{rapp-wambach2000,brown-rho2002}). A simple superposition of pp collisions cannot explain the data and new physics
has been invoked. The $\pi^+\pi^-$ annihilation channel, ($\pi^+\pi^-\rightarrow \rho \rightarrow e^+e^-$) accounts for
a fraction of the enhancement (dashed line) and provides first evidence for the thermal radiation emitted from a
dense hadron gas. However, this process is insufficient for a quantitative description of the data. In order to do that,
it is necessary to introduce in-medium modification of the intermediate $\rho$ meson. Two main venues have been used: (i) a
decrease of the $\rho$-meson mass in the dense fireball \cite{li-ko-brown} as a precursor of CSR, following the original
Brown-Rho scaling \cite{brown-rho} (dash-dotted line). In this scenario, the $\rho$-meson mass scales with the quark condensate
$<\overline{q}q>$ and the latter drops due to the high baryon density and (ii) a broadening of the $\rho$-meson spectral function 
resulting from the scattering of the $\rho$ meson mainly off the baryons in the dense hadronic medium 
 \cite{rapp-wambach2000} (solid line). Both approaches rely on the high baryon
density at mid-rapidity which, at CERN energies, mainly originates from baryon stopping, and both achieve good
agreement with the CERES data in the mass region m = 0.2-0.6~GeV/c2 (see Fig.~\ref{fig:ceres-95-96}). A similar or even 
\begin{figure}
  \vspace{-0.7cm}
  \resizebox{0.50\textwidth}{!} {\includegraphics{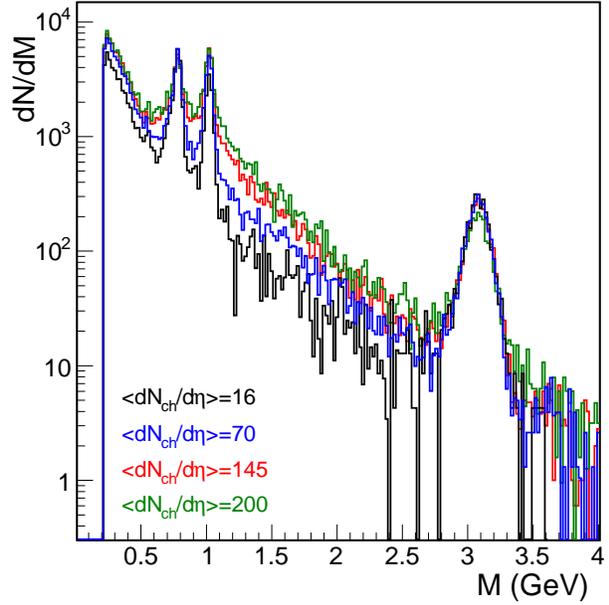} }
  \caption{Invariant mass spectra for four bins of centrality in In+In collisions at 158~AGeV, from NA60 \cite{na60}.}
\vspace{-0.7cm}
\label{fig:na60}
\end{figure} 
stronger enhancement has been observed by CERES in Pb+Au collisions at 40~AGeV and this
enhancement is also equally well reproduced by the two models \cite{ceres-pb40}. The success of these two different
approaches, one relying on quark degrees of freedom and the other one based on a pure hadronic model, has attracted
much debate raising also the interesting possibility of quark-hadron duality at these relatively low-masses
\cite{rapp-duality}. Precise data with high mass resolution and high statistics will be very valuable. In the mass 
region m = 0.2-0.6~GeV/c$^2$, the two models are so similar that it will be very difficult 
to distinguish between them. However, at higher masses, in the region between the $\omega$ and the $\phi$, the two models 
differ markedly as shown in Fig.~\ref{fig:ceres-95-96} and the difference becomes even more pronounced with higher mass resolution.
High statistics and high resolution data should therefore be able to discriminate between the two models. The CERES data
obtained with the upgraded spectrometer (not yet available in a final form) seem marginal for that. The NA60 experiment has the 
potential to achieve this goal.
 
\begin{figure*}
  \resizebox{0.9\textwidth}{!} {
  \hspace{4.0cm} \includegraphics{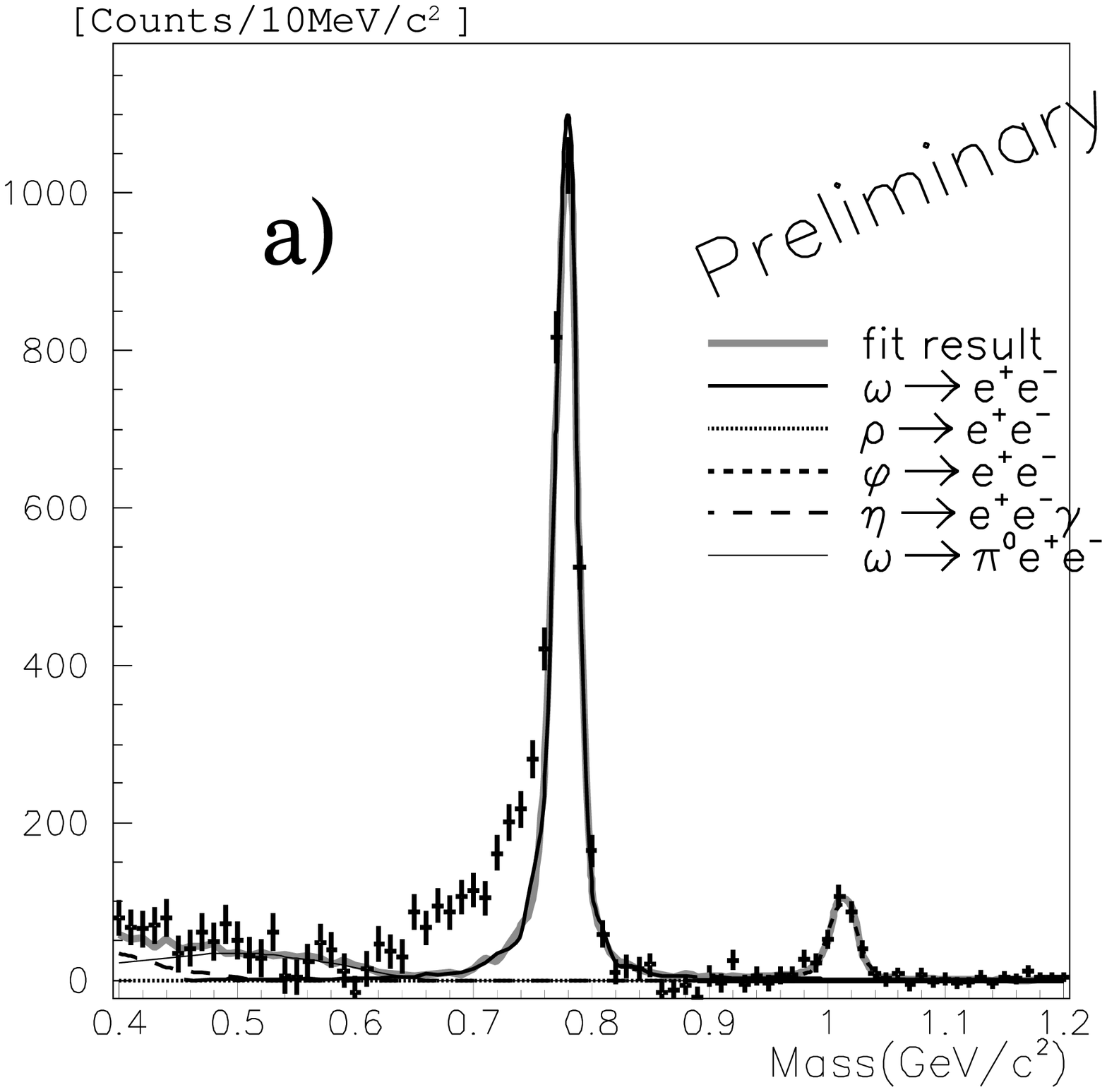}  \hspace{6.0cm}
  \includegraphics{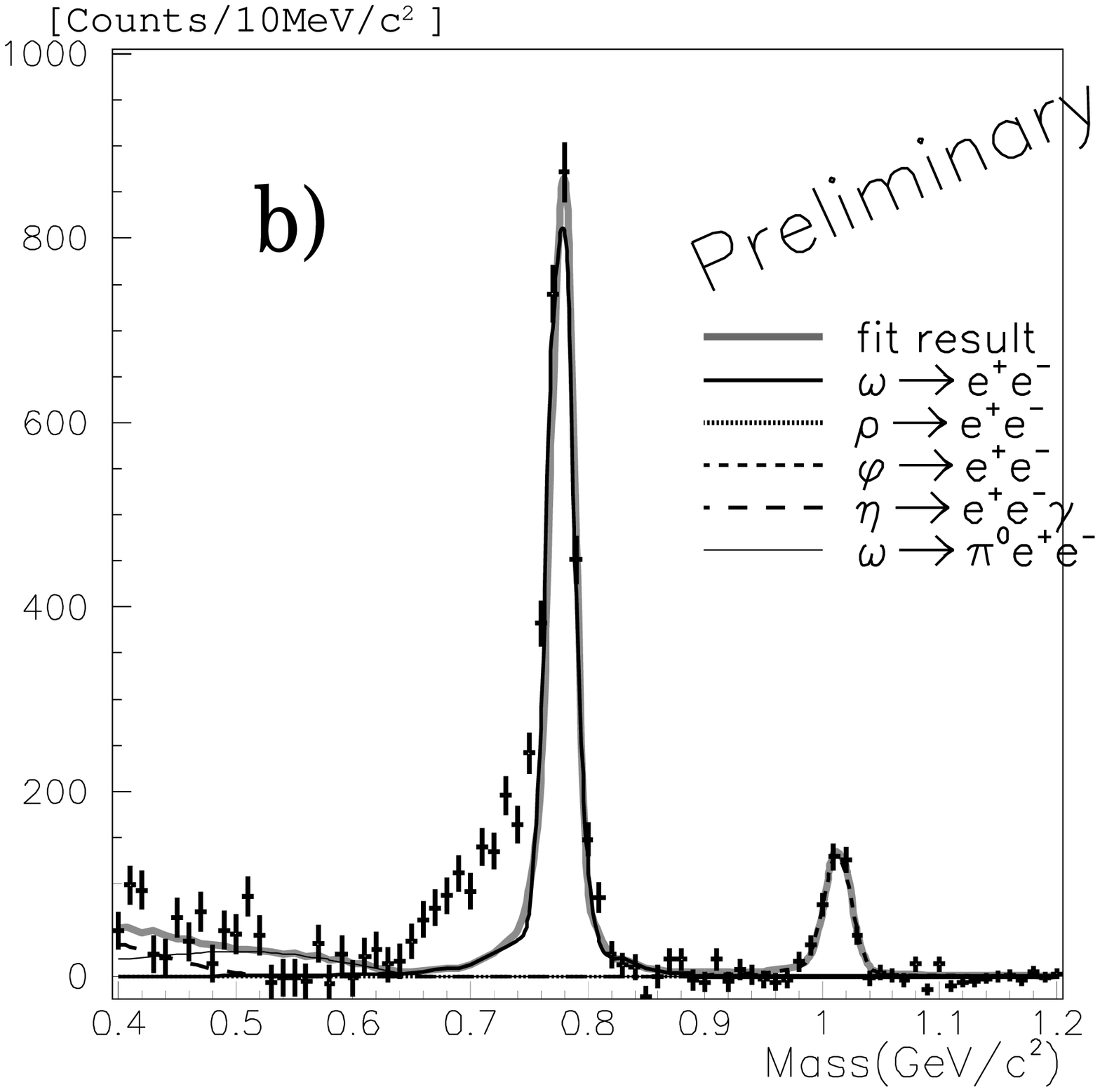}}
  \caption{Invariant mass spectra, after combinatorial background subtraction, measured in 12~GeV p+C (a) and p+Cu (b)
           collisions by the E325 experiment at KEK \cite{kek}.}
  \label{fig:kek}
\end{figure*}
\begin{figure*}
  \resizebox{0.9\textwidth}{!} {
  \hspace{2.0cm} \includegraphics{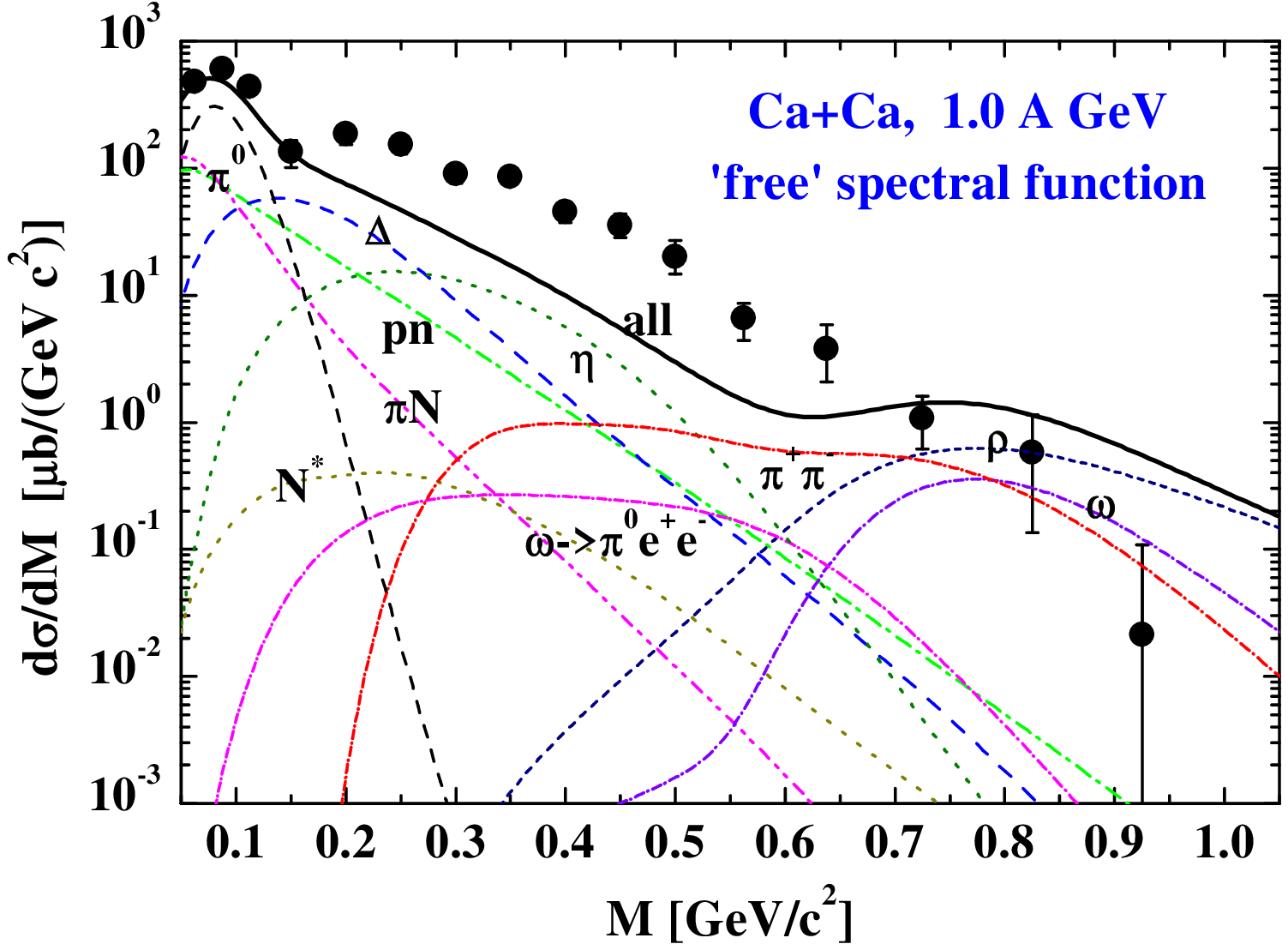} \hspace{2.0cm}
  \includegraphics{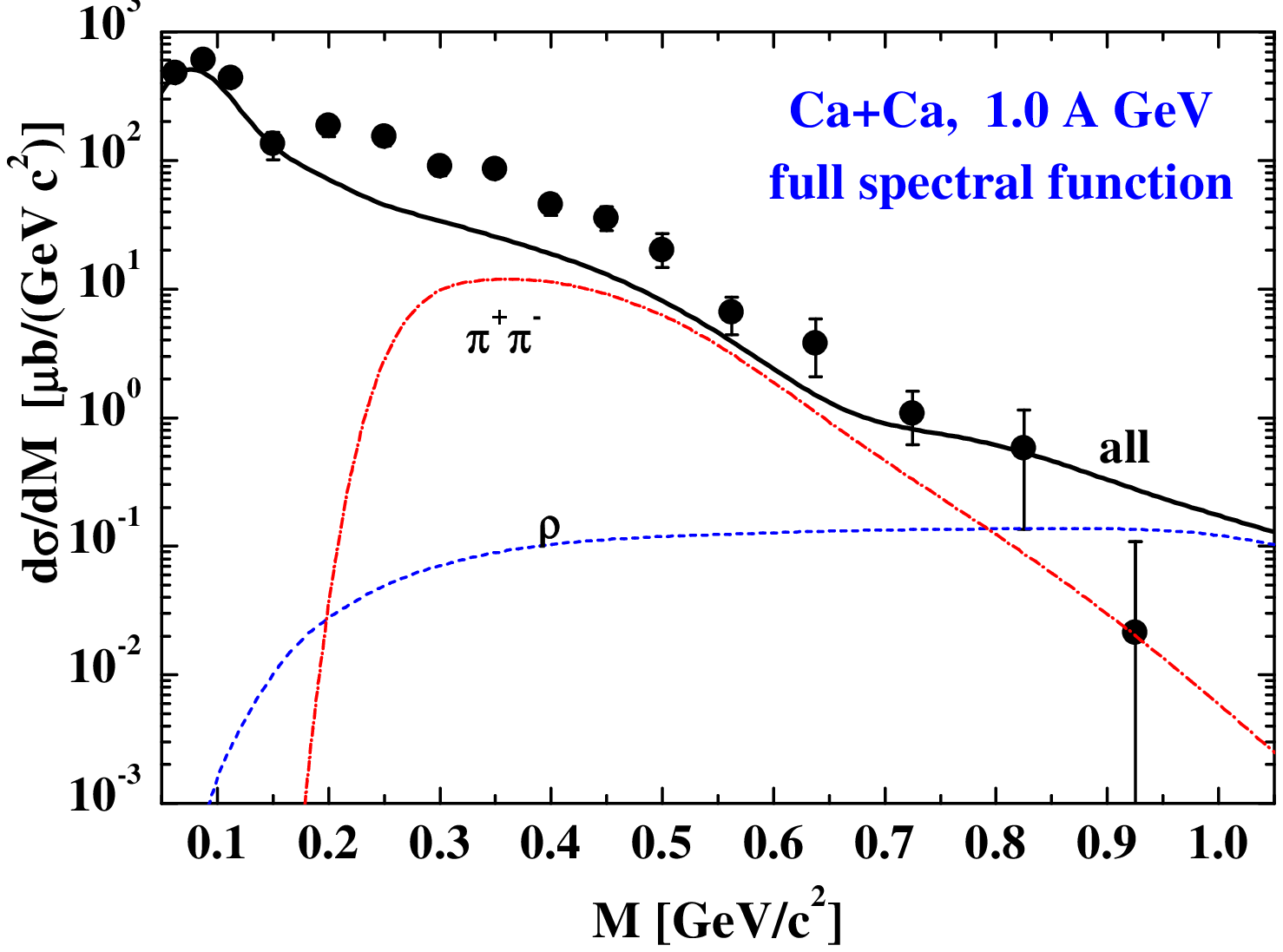}}
  \caption{DLS results in Ca+Ca collisions at 1~AGeV compared to calculations from ref.~\cite{bratkovskaya}. The left panel
  uses the free $\rho$ meson spectral function whereas the right panel shows the calculations with the full $\rho$ spectral function}
  \label{fig:dls}
\end{figure*} 

NA60 has measured low-mass di-muon production in In+In collisions at 158~AGeV. Raw spectra at four centralities are shown 
in Fig.~\ref{fig:na60}, arbitrarily normalized at the $\omega$ peak \cite{na60}. The experiment has limited p$_T$ acceptance at low-masses 
but it has an excellent mass resolution as illustrated in the figure. The $\omega$ and $\phi$ peaks are clearly resolved 
with a mass resolution of 23 MeV/c$^2$ at the $\phi$. NA60 is also  a high luminosity experiment. The spectra in Fig.~\ref{fig:na60} 
represent approximately 1/3 of the whole data sample. With such high quality data NA60 should be able to provide significant information
on the low-mass dilepton enhancement.

\subsection{Results at low energies}
\label{subsec:low_energies}
Low-mass electron pairs have also been measured in p+A collisions at 12~GeV at KEK by the E325 experiment
\cite{kek}. The measurements were performed in the target rapidity region to enhance the decay probability of the
virtual photon inside cold nuclear matter. An essential feature of this experiment is its excellent mass resolution,
slightly better than 1\% at the $\phi$ mass, which made it possible to observe a small but significant difference in
the invariant mass spectrum below the $\omega$ meson, in p+C and p+Cu collisions, see Fig.~\ref{fig:kek} \cite{kek}.
More recently, the same experiment reported the observation of similar differences below the $\phi$ meson
\cite{kek2}. Both effects are attributed to modifications of the meson spectral shape in nuclear matter at normal
density as predicted e.g. by the models of Brown and Rho \cite{brown-rho} or Hatsuda and Lee \cite{hatsuda-lee}. Further 
information in this energy regime, including nucleus-nucleus collisions, will have to await the new experiment CBM 
being designed at the future FAIR facility of the GSI.

At even lower energies, the DLS results remain a puzzle. The DLS Collaboration has measured low-mass $e^+e^-$  
pairs in 1 AGeV Ca+Ca collisions at the BEVALAC \cite{dls97}. The Ca results are displayed in Fig.~\ref{fig:dls}. In the 
left panel, the results are compared to a cocktail including hadrons decays, proton-neutron (pn) and pion-nucleon ($\pi$N) 
bremsstrahlung and also the pion annihilation ($\pi^+\pi^- \rightarrow \rho \rightarrow e^+e^-$) using the vacuum 
or free $\rho$ spectral function \cite{bratkovskaya}. An enhancement of low-mass pairs is observed in the mass range 
m = 0.2 - 0.6~GeV/c$^2$
which looks qualitatively similar to the CERES results. However, attempts to reproduce this enhancement have so far failed. 
The right panel in Fig.~\ref{fig:dls} compares the data with the same cocktail using this time the $\rho$ meson spectral
function modified in the medium. Part of the enhancement is explained but the calculations are still a factor 
of 2-3 below the data \cite{bratkovskaya}. Over the last few years, the HADES experiment at the GSI has collected data on 
di-electron production in C+C collisions at 1 and 2~AGeV and first results should be available soon.

\subsection{Prospects at RHIC}
\label{subsec :RHIC}
\begin{figure}
  \resizebox{0.50\textwidth}{!} 
  {\includegraphics{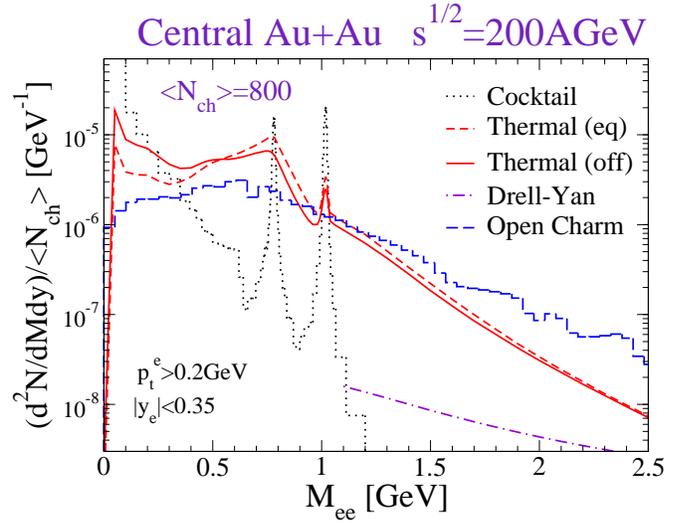} } 
  \caption{Invariant mass spectrum of e+e- pairs emitted in Au+Au collisions at $\sqrt{s_{NN}}$ =200 Gev calculated in 
           the PHENIX acceptance comparing  in-medium modified spectrum (thermal) to the cocktail of hadron decays, open 
           charm decays and Drell-Yan annihilation \cite{rapp-rhic}. }
  \label{fig:rapp-rhic}
\end{figure}

   The study of low-mass electron pairs under the much better conditions offered at RHIC -higher
initial temperature, larger energy density, possibly larger volume and longer lifetime of the system- promises to be
very interesting. The total baryon density, which is the key factor responsible for in-medium modifications of the
$\rho$ meson at SPS energies both in the dropping mass and the collision broadening scenarios, is almost as high at
RHIC as at SPS, contrary to previous expectations.  At SPS, most of the baryons at mid-rapidity are participant
nucleons. At RHIC there is a strong decrease in nuclear stopping but this is compensated by a copious production of
baryon-antibaryon pairs such that the total baryon density is accidentally the same in both cases. Furthermore, the
temperature factor which played a minor role at SPS energies is expected to be much more significant at RHIC energies.
Updated calculations that incorporate results from global and hadronic observables at RHIC, predict
indeed that the enhancement of low-mass electron pairs persists at the collider with at least comparable strength as
shown in Fig.~\ref{fig:rapp-rhic} \cite{rapp-rhic}. The figure compares the hadronic cocktail with the in-medium modified spectrum
(labeled "thermal" in the figure). In addition to the $\rho$, the calculations also predict strong in-medium
modifications of the $\omega$ and $\phi$ mesons. These are much less dramatic than in the case of the $\rho$ meson but
should nevertheless be readily observable with  a mass resolution comparable or better than the natural width of the
resonances. The figure shows that the low-mass region will also be sensitive to open charm production.

\begin{figure}
  \resizebox{0.50\textwidth}{!} {\includegraphics{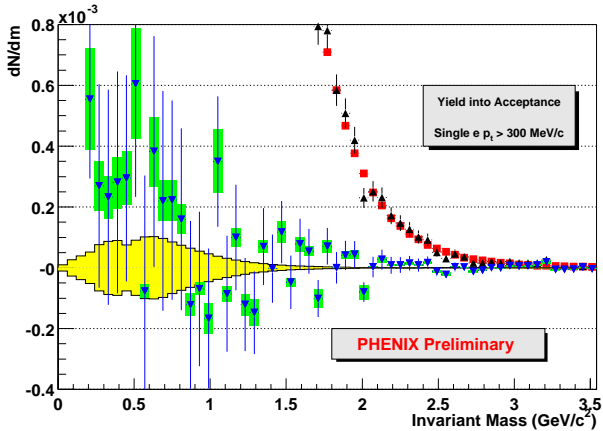} }
  \caption{Low-mass dilepton mass spectrum from analysis of Au+Au data after subtraction of the 
           combinatorial background from mixed events with an electron from one event and positron from a different event 
           \cite{phenix-low-masses}.}
\label{fig:phenix-low-masses}
\end{figure}

PHENIX is the only experiment at RHIC that has the potential to measure low-mass electron pairs. The mid-rapidity spectrometers
have good electron identification capabilities by combining a RICH detector with an electromagnetic calorimeter \cite{phenix-rich}. 
However, the strong magnetic field, starting at the collision vertex, causes a limited acceptance of tracks originating from
$\pi^0$ Dalitz decays and $\gamma$ conversions which leads to an overwhelming yield of combinatorial background pairs.
 Fig.~\ref{fig:phenix-low-masses} shows the electron-positron 
invariant mass spectrum from an analysis of the 2002 Au+Au data at $\sqrt{s_{NN}}=200$~GeV, after subtraction of the 
combinatorial background determined with a mixed event technique \cite{phenix-low-masses}. The signal to background ratio is of the order of S/B 
$\sim$1/200 (1/500) with a single electron p$_T$ cut of 300 (200) MeV/c making the measurement of the low-mass pair 
continuum practically impossible.

PHENIX has developed a Hadron Blind Detector (HBD) as an upgrade to overcome this problem. The main task of the HBD is to recognize and reject 
electron tracks originating from $\pi^0$ Dalitz decays and $\gamma$ conversions. The strategy is to exploit the fact that the 
opening  angle of electron pairs from these sources is very small compared to pairs of heavier mass. The HBD is therefore located 
in a field-free region where the pair opening angle is preserved. An inner coil, recently installed in the central arms of PHENIX, 
counteracts the main field of the outer coils creating an almost field-free region close to the vertex and extending to $\sim$50-60~cm 
in the radial direction. Conceptual Monte Carlo simulations show that with an opening angle
cut of $\sim$200~mrad, the combinatorial background arising from $\pi^0$ Dalitz decays and  $\gamma$ conversions is
suppressed by approximately two orders of magnitude while preserving $\sim$50\% of the signal \cite{TN391}.

The HBD is a windowless Cherenkov detector operated with pure CF$_4$, in a proximity focus configuration. The detector
consists of a 50~cm long radiator directly coupled to a triple GEM detector \cite{sauli} which has a CsI photocathode
evaporated on the top face of the first GEM foil and a pad readout at the bottom of the GEM stack. In this scheme the
Cherenkov light from particles passing through the radiator is directly collected on the photocathode forming a
circular blob image rather than a ring as in a RICH detector. The validity of this novel HBD concept has been 
demonstrated in a comprehensive R\&D program \cite{hbd1,hbd2} paving the way for the incorporation of such a detector 
in the PHENIX experiment.

\section{$\phi$ meson production}
\label{sec:phi}
 The $\phi$  meson is considered a sensitive probe for studying in-medium modifications of the vector mesons spectral
shape (mass and/or width) as chiral symmetry restoration is approached \cite{pal-ko-lin}. With a lifetime of 
$\tau$ = 46~fm/c the $\phi$ mesons predominantly decay outside the medium after regaining the vacuum properties, with only 
a small fraction decaying inside the medium. 
Since the measurement integrates over the collision's history this may result 
in a small modification of the $\phi$ meson line shape manifesting itself in a tail at lower masses. Nevertheless,
PHENIX and NA60 with their excellent mass resolution might be able to observe such modifications. 
Furthermore, the simultaneous measurement of the $\phi$ decay into $K^+K^-$ and dileptons provides a very powerful tool in the
search for such in-medium modifications. Since the $\phi$-meson mass is close to twice the kaon mass, even a small 
decrease of the $\phi$ meson mass may induce a large decrease in the $\phi \rightarrow K^+K^-$ yield. The interpretation 
might not be straightforward since the scattering and absorption of low-momentum kaons in the medium can also lead to a 
smaller yield of $\phi$ mesons reconstructed from the kaon channel than that reconstructed from the dilepton channel.  
   
   There is only limited and even controversial information about the production of the vector mesons $\omega$ and $\phi$
at SPS energies. The $\phi$ meson production in central collisions at 158 AGeV has been studied via the muon decay channel 
by NA50 (Pb+Pb) \cite{na50-phi} and  via the kaon 
decay channel by NA49 (Pb+Pb) \cite{na49-phi} and CERES (Pb+Au)\cite{ceres-sqm04}. The NA50 experiment observed a 
significant increase of the $\phi/(\rho+\omega)$ ratio with centrality which is attributed to an enhancement of the 
$\phi$ production \cite{pal-ko-lin}. The NA49 experiment found an increase of the $\phi/\pi$ ratio which is also attributed to
an enhancement of the $\phi$ production. However, the inverse slope parameters T, as obtained from an exponential fit to
the $\phi$ m$_T$ distributions are T=228$\pm$10~MeV, 283$\pm$11~MeV and 305$\pm$15~MeV in the NA50, CERES and NA49 experiments, 
respectively. There is an even larger discrepancy in their measured $\phi$ yields. The NA50 yields are larger than those of 
NA49 by factors of 2 to 4 in the common m$_T$ range covered by the two experiments. It is unclear whether these
differences are of experimental origin or a manifestation of in-medium effects. Additional insight on this issue is provided
by the first NA60 results from In+In collisions. The NA60 $\phi/\omega$ ratio is in very good agreement with the NA50 results.
However, the temperature parameter T that shows a clear increase with centrality, appears closer to the NA49 
than the NA50 results \cite{na60}.

First results on $\phi$ meson production at RHIC are already available. Using the run-2 data, PHENIX attempted an 
analysis of $\phi$  meson production through the $K^+K^-$ and $e^+e^-$ decay channels \cite{phenix-phi-qm02}. The 
study of the $\phi$  meson production through the $K^+K^-$ decay channel has recently been completed \cite{phenix-phi}. 
The results include the centrality dependence of particle density dN/dy, particle spectra and inverse slope parameter 
T. A line shape analysis revealed no significant change in the centroid and width values of the $\phi$ meson from 
the PDG accepted values. The statistics accumulated in run-2 were insufficient for a similar quality study of meson 
production through the $e^+e^-$ decay channel. This crucial comparison study is not yet available and is expected to 
be done with the higher statistics data accumulated in the fourth run of RHIC (2004).

\section{Thermal photons}
\label{sec:photons}
A central topic of interest in the field of relativistic heavy-ion collisions is the identification of thermal 
radiation emitted by the dense medium. Such radiation is a direct fingerprint of the matter formed, the quark-gluon 
plasma (QGP) or a high-density hadron gas (HG). The elementary processes involved are well known. The main channels 
for the production of
thermal photons are $q\overline{q}$ annihilation and QCD Compton scattering $qg \rightarrow q\gamma$ in the QGP phase and pion
annihilation ($\pi^+\pi^-\rightarrow \rho\gamma$ or $\eta\gamma$) and hadron-hadron scattering (like $\pi\rho
\rightarrow \pi\gamma$) in the HG phase. The unambiguous identification of the thermal radiation from the QGP is
considered as a very strong signal of deconfinement and its spectral shape should provide a direct measurement of the
plasma temperature. The absolute yields, obtained by integrating the emission rate over the space-time evolution of the
collision, have been calculated by several authors. The results point to rather well established features. At
the same temperature, thermal photons from partonic and hadronic processes have a similar production rate. Their overall yield is
relatively small compared to the inclusive photon yield.  In a recent elaborate calculation 
the transverse momentum range p$_T$ = 1-3~GeV/c appears as the most promising window where the QGP radiation could
dominate over other contributions in central Au+Au collisions at $\sqrt{s_{NN}}$= 200 GeV \cite{turbide}.

As mentioned in Section \ref{sec:low-mass}, the excess of low-mass electron pairs provide first evidence of thermal radiation
from a high density hadronic system at the SPS. However, there is no conclusive evidence for QGP thermal photons at the SPS. 
The initial measurements of WA80 \cite{wa80-photons} and CERES \cite{ceres-photons} with the 200 AGeV S beam yielded 
only upper limits of ~15\% of the total photon yield. In the two experiments the sensitivity was actually limited not by 
the statistics but rather by the
systematic errors, too large to identify a thermal photon source which may only be of the order of a few percent of the
inclusive photon yield. The WA98 experiment observed a small photon excess in central Pb+Pb collisions at 158~AGeV,
whereas in peripheral collisions the yield is consistent with the expectations from hadron decays \cite{wa98-photons}.
The excess is only a 1-2$\sigma$ effect and occurs only at p$_T >$ 1.5 GeV/c. These results have attracted considerable
theoretical interest. Although it is possible to explain the excess by the contribution from direct photons including
k$_T$ broadening (Cronin effect) it is not possible to rule out a small contribution from QGP thermal photons \cite{turbide}.

\begin{figure}
  \resizebox{0.50\textwidth}{!} {\includegraphics{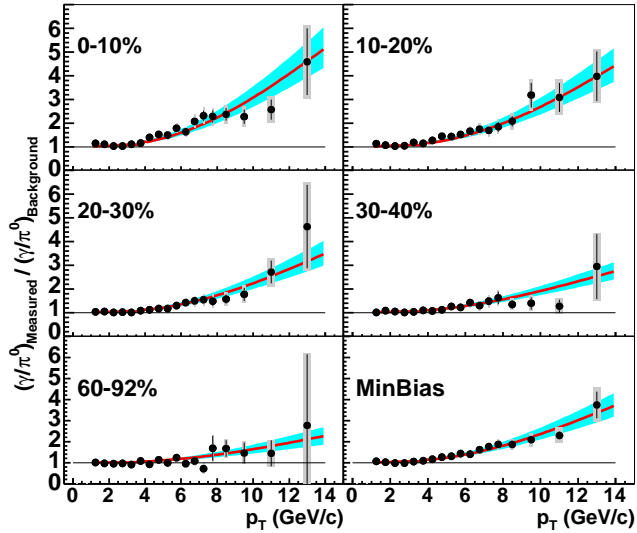} }
  \caption{PHENIX measurement of real photons per $\pi^0$
          relative to the calculated background from hadron decays in central Au+Au collisions at 
          $\sqrt{s_{NN}}$ = 200 GeV. The curves represent NLO pQCD calculations of direct photons in p+p collisions 
          scaled to Au+Au by the number of collisions N$_{coll}$ for each centrality selection \cite{phenix-photons}. }
  \label{fig:phenix-photons}
\end{figure}

   A first measurement of real photons in Au+Au collisions at $\sqrt{s_{NN}}$ = 200~GeV was performed by PHENIX using
data from the second RHIC run, in the year 2002 \cite{phenix-photons}. After subtracting the hadron decays, there is clear 
evidence of an excess which can be explained by direct photons from initial hard scattering using NLO pQCD calculations 
for p+p collisions scaled to Au+Au by the average number of collisions for each centrality bin 
(see Fig.~\ref{fig:phenix-photons}). 
The errors however, are relatively large leaving room for a comparable contribution of thermal photons. The high statistics 
accumulated in the fourth RHIC run shall provide the first real opportunity to search for the QGP thermal photons in PHENIX.

\section{Summary}
\label{sec:summary}
The low-mass dilepton experiments, covering a very broad energy range from the BEVALAC to the SPS energies, 
have produced very exciting and stimulating results, most of them not fully understood and with a possible link to chiral 
symmetry restoration. A new generation of experiments with considerably improved performance will soon start to 
provide new results on the same energy range (HADES and NA60). The extension of these measurements to RHIC energies
by the PHENIX experiment promises also to be very interesting. The quest for the thermal radiation from the QGP remains difficult. Results of
the first real search for this elusive signal at RHIC using the high statistics Au+Au data from run-4 should 
be available soon.

%

\end{document}